  \documentclass[12pt]{aastex7}

\hypersetup{linkcolor=red,citecolor=blue,filecolor=cyan,urlcolor=magenta}
\usepackage{verbatim}  % Add this in your preamble

\newcommand{\Lya}     {Ly$\alpha$}    %  Lyman alpha
\newcommand {\HI}        {\ion{H}{1}}   %  HI
\newcommand {\HeI}     {\ion{He}{1}}   %  HeI
\newcommand {\HeII}     {\ion{He}{2}}  %  HeII

\newcommand {\kms}    {km~s$^{-1}$}

\newcommand {\etal}   {et~al.}

\shorttitle{Beta CMa Photoionization}
\shortauthors{Shull et al.}

\begin{document}

\title{Beta Canis Majoris:  The Other Major Ionization Source \\
of the Local Interstellar Clouds } 

\author[0000-0002-4594-9936] {J. Michael Shull} {\email{michael.shull@colorado.edu} }
\affiliation{Department of Astrophysical \& Planetary Sciences, University of Colorado, 
     Boulder, CO 80309, USA;  michael.shull@colorado.edu }
\affiliation{Department of Physics and Astronomy, University of North Carolina, Chapel Hill, NC 27599, USA; }

\author{Rachel M. Curran} \email{rcurran@unc.edu}
\affiliation{Department of Physics \& Astronomy, University of North Carolina, Chapel Hill, NC 27599 USA;}

\author[0000-0001-8426-1141] {Michael W. Topping} \email{michaeltopping@arizona.edu}
\affiliation{Steward Observatory, University of Arizona, Tucson, AZ 85721, USA}

\email{michael.shull@colorado.edu, rcurran@unc.edu, michaeltopping@arizona.edu} 

%%%%%%%%%%%%%%%%%%%%%%%%%%%%%%%%%%%%%%%%%%

%%%%%%%%%%%%%%%%%%%%%%%%%%%%%%%%%%%%%%%%
%%
%%           ABSTRACT
%%
%%%%%%%%%%%%%%%%%%%%%%%%%%%%%%%%%%%%%%%%

\begin{abstract}

Two nearby B-type stars, $\epsilon$~CMa ($124\pm2$~pc) and $\beta$~CMa ($151\pm5$~pc),
are important contributors to the photoionization of the local interstellar cloud (LIC).  At spectral
type B1~II-III, $\beta$~CMa is slightly hotter than $\epsilon$~CMa (B2~II-III), but its ionizing flux 
at Earth is attenuated by a much larger \HI\ column density.  At the external surface of the LIC, 
the two stars produce similar fluxes in the Lyman continuum (LyC).  From the $\beta$~CMa 
angular diameter, bolometric flux, and position on the Hertzsprung-Russell diagram, 
we obtain a consistent set of stellar  parameters:  $T_{\rm eff} = 25,180\pm1120$~K, 
$\log g = 3.70\pm0.08$, radius $R = 8.44\pm0.56~R_{\odot}$, mass $M = 13\pm1~M_{\odot}$, 
and luminosity $L = 10^{4.41\pm0.06}~L_{\odot}$.   The EUVE-observed fluxes and non-LTE 
model atmospheres are used to determine the ionizing photon production rate 
$Q_{\rm H} = 10^{46.0\pm0.1}~{\rm s}^{-1}$ and fluxes incident on the local clouds, 
$\Phi_{\rm HI} \approx 3700~{\rm cm}^{-2}~{\rm s}^{-1}$ and
$\Phi_{\rm HeI} \approx 110~{\rm cm}^{-2}~{\rm s}^{-1}$ in the \HI\ and \HeI\ continua.  
The corresponding photoionization rates are  $\Gamma_{\rm HI} \approx 1.5\times10^{-14}$~s$^{-1}$ 
and $\Gamma_{\rm HeI} \approx 7.3\times10^{-16}$~s$^{-1}$.  Within the local cloud, the LyC flux 
is attenuated by an \HI\ column density $N_{\rm HI} = (1.9\pm0.1)\times10^{18}~{\rm cm}^{-2}$,
with optical depth $\tau_{\rm LL} = 12.0\pm 0.6$ at the Lyman limit.  The radial velocities and proper 
motions of $\beta$~CMa and $\epsilon$~CMa indicate that both stars passed within $10\pm1$~pc 
of the Sun approximately 4.4~Myr ago, with incident ionizing fluxes 180--200 times larger.  Their 
EUV radiation photoionized and heated the tunnel in the local interstellar gas, associated 
dynamically with past supernova explosions in the Sco-Cen OB association.

\end{abstract}

\section{\bf Introduction}  

\bigskip

The B-type giant star Beta~Canis Majoris ($\beta$~CMa), also known as HD~44743 and Mirzam, 
is one of several bright sources of extreme ultraviolet (EUV) radiation  (J.\ Dupuis \etal\ 1995; 
J.\ Vallerga \& B.\ Welsh 1995; N.\ Craig \etal\ 1997) observed by the Extreme Ultraviolet Explorer 
(EUVE).  Both $\epsilon$~CMa and $\beta$~CMa  are located in the direction ($\ell = 232\pm5$, 
$b = -15\pm4)$ of a low-density, highly ionized tunnel in the interstellar medium  (P.\ Frisch \& 
D.\ York 1983; C.\ Gry \etal\ 1985; B.\ Welsh 1991; J.\ Vallerga 1998; D.\  Sfeir \etal\ 1999;
J.\ Linsky \etal\ 2019).  Both stars contribute to the \HI\ photoionization rate of the Local Interstellar 
Cloud or LIC  (J.\ Vallerga 1998).  A portion of the $\beta$~CMa ionizing spectrum was measured 
by EUVE between 504--720~\AA.   However, any flux between 750--912~\AA\ would have been 
highly attenuated by the interstellar medium (ISM).  After correcting for ISM absorption, we find
that $\beta$~CMa and $\epsilon$~CMa have comparable fluxes in the Lyman continuum (LyC) 
incident on the surface of the local clouds.  

\medskip

In our previous paper (J. M.\ Shull \etal\ 2025) we analyzed the stellar parameters of $\epsilon$~CMa, 
and derived the intervening \HI\ column density and the \HI\ photoionization rate.  In this paper, we
provide similar calculations of the stellar parameters of $\beta$~CMa, the ISM column density, and 
photoionization rates of \HI\ and \HeI.   We employ non-LTE, line-blanketed model atmospheres to 
anchor the relative fluxes in the stellar EUV continuum shortward of the ionization edges of  
\HI\ ($\lambda \leq 911.75$~\AA) and \HeI\ ($\lambda \leq 504.26$~\AA).  We then use observed
fluxes in the EUV (504--720~\AA) and far-UV (1000--1400~\AA) to determine the attenuation of 
the LyC radiation in both the stellar atmosphere and ISM.  

\medskip

In Section~2, we derive a consistent set of stellar parameters (radius, mass, effective temperature, 
luminosity) based on measurements of the parallax distance, stellar angular diameter, and total flux.   
In Section~3, we analyze the EUV flux attenuation in the stellar atmosphere and derive an interstellar 
\HI\ column density, $N_{\rm HI} = (1.9\pm0.1)\times10^{18}~{\rm cm}^{-2}$, corresponding to an \HI\ 
optical depth $\tau_{\rm LL} = 12.0\pm0.6$ at the Lyman limit (LL).   The non-LTE stellar model 
atmosphere predicts a flux decrement factor of $\Delta_{\rm LL} = 67\pm7$ at 912~\AA\ but shows
little absorption at the \HeI\ ionization edge (504~\AA).  This is evidently a non-LTE effect of backwarming 
from the line-blanketed B-star wind.  
In Section~4  we discuss the implications of our results for the ionization structure of the local interstellar 
clouds.  We find that $\epsilon$~CMa and $\beta$~CMa have comparable photoionization rates of \HI\ at 
the external surface of the local clouds.  However, $\beta$~CMa has a much larger ionization rate of \HeI,
owing to its higher effective temperature $T_{\rm eff} \approx 25,000$~K, compared to 21,000~K for
$\epsilon$~CMa.  

\bigskip

\section{\bf Revised Properties of Beta Canis Majoris}

\bigskip

\subsection{Basic Stellar Parameters} 

\bigskip

In a classic survey of southern B-type stars, J.\ Lesh (1968) listed the spectral type (SpT) of
$\beta$~CMa as B1~II-III, with apparent and absolute visual magnitudes $(m_V = 1.97$, 
$M_V = -4.57$) and a spectrophotometric distance of 203~pc.  A SpT of B1~II-III was 
confirmed by I.\ Negueruela \etal\ (2024).  However, the {\it Hipparcos} parallax measurement of 
$6.62\pm0.22$~mas (F.\ van Leeuwen 2007) provided a shorter distance $d = 151\pm5$~pc, 
with a modulus ($m_V$--$M_V) = 5.895 \pm0.071$ smaller by 0.64 mag.  
In other papers, the distance to $\beta$~CMa was quoted variously as 213~pc (R.\ Bohlin 1975), 
206~pc (B.\ Savage \etal\ 1977; R.\ Bohlin \etal\ 1978; J.\  Cassinelli \etal\ 1996), and 203~pc 
(B.\ Welsh 1991; J. M.\ Shull \&  M.\ Van Steenberg 1985).  The new distance modulus implies 
an absolute magnitude $M_V = -3.925 \pm 0.071$, similar to the value of $-4.0$ given in 
J.\ Lesh (1968) for B1~III spectral type.  

\medskip

A key physical measurement for $\beta$~CMa is its angular diameter $\theta_d = 0.52\pm0.03$~mas 
(R.\ Hanbury Brown \etal\ 1974) found with the stellar interferometer at the Narrabri Observatory.  
From this and a parallax distance $d = 151\pm5$~pc, we derive a stellar radius of
\begin{equation}
     R = \left( \frac {\theta_d \, d} {2} \right) = 5.87 \times10^{11}~{\rm cm}  
           ~~ (8.44\pm0.56~R_{\odot})   \; , 
\end{equation}  
where we combined the relative errors on $\theta_d$ (5.8\%) and $d$ (3.3\%) in quadrature.  
The revision in the $\beta$~CMa distance from 203~pc to 151~pc also changes its inferred
mass and surface gravity.   L.\ Fossati \etal\ (2015) used the new distance $d = 151\pm5$~pc 
and evaluated the radius and mass from two sets of evolutionary models:  
$R = 7.4^{+0.8}_{-0.9}~R_{\odot}$ and $M = 12.0^{+0.3}_{-0.7}~M_{\odot}$ (tracks from 
C.\ Georgy \etal\ 2013) and $R = 8.2^{+0.6}_{-0.5}~R_{\odot}$ and 
$M = 12.6^{+0.4}_{-0.5}~M_{\odot}$ (tracks from L.\ Brott \etal\ 2011).  These radii are in good
agreement with our radius ($8.44\pm0.56~R_{\odot}$) and the $13\pm1~M_{\odot}$
gravitational and evolutionary masses that we derive below. 

\medskip

The absolute magnitude and bolometric magnitude of $\beta$~CMa need revision based on
its new distance $d = 151\pm5$~pc, but the stellar classification has remained consistent at 
B1~II-III  (J.\ Lesh 1968; L.\ Fossati \etal\ 2015; I.\ Negueruela \etal\ 2024).  At this SpT, the
absolute magnitude $M_V = -4.0$ (J.~Lesh 1968 value at B1.5~III), and  we adopt a bolometric 
correction $(M_{\rm bol}$--$M_V) \approx -2.5$ from Figure~5  of M.~Pedersen \etal\ (2020) at 
$T_{\rm eff} = 25,000$~K and $\log g$ = 3.5--4.0.  This yields a bolometric absolute magnitude 
$M_{\rm bol} = -6.425\pm0.07$ and luminosity of $10^{4.47}~L_{\odot}$, based on the solar 
bolometric absolute magnitude $M_{{\rm bol},\odot} = 4.74$.  This luminosity comes with some
uncertainty from the applied bolometric correction.   

\bigskip 

Values of $R$, $T_{\rm eff}$, $g$, and $L$ must satisfy the relations $g =  (G M/R^2)$ and 
$L = 4 \pi R^2 \sigma_{\rm SB} T_{\rm eff}^4$, where 
$\sigma_{\rm SB} = 5.6705 \times 10^{-5}$ erg~cm$^{-2}$~s$^{-1}$~K$^{-4}$ is the 
Stefan-Boltzmann constant and $R = (\theta_d \, d/2)$.  Here, $\theta_d = 2.52 \times10^{-9}$~rad 
($0.52\pm0.03$ mas) is the measured stellar angular diameter (R.\ Hanbury Brown \etal\ 1974).  
The integrated stellar flux is $f = (36.2 \pm 4.9) \times10^{-6}~{\rm erg~cm}^{-2}~{\rm s}^{-1}$ 
from A.\ Code \etal\ (1976), who combined ground-based visual and near-infrared photometry 
with ultraviolet observations (1100--3500~\AA) taken by the Orbiting Astronomical Observatory 
(OAO-2).  This flux included an extrapolated value for shorter wavelengths 
$f(\lambda \leq 1100~{\rm \AA}) = (7.5\pm3.8)\times10^{-6}~{\rm erg~cm}^{-2}~{\rm s}^{-1}$.
Even after correcting the observed EUV flux for interstellar absorption, the Lyman continuum 
represents less than 1\% of the bolometric flux.  The constraints from stellar angular diameter 
and total flux give a useful relation for the bolometric effective temperature 
$T_{\rm eff} = (4f / \sigma_{\rm SB} \, \theta_d^2)^{1/4}$,
\begin{equation} 
    T_{\rm eff} = (25,180\pm1120~{\rm K}) \left( \frac {f} {36.2\times10^{-6}} \right)^{1/4} 
             \left( \frac {\theta_d} {0.52~{\rm mas}} \right)^{-1/2}       \;  . 
\end{equation}
The stellar radius and bolometric relation with $T_{\rm eff}$ suggest a total luminosity,
\begin{equation}
   L \approx (25,800\pm3900~L_{\odot}) \left( \frac {T_{\rm eff}}{25,180~{\rm K}} \right)^4 
          \left( \frac {R}{8.44~R_{\odot}} \right)^2  \; .
\end{equation}
With the relative errors on parallax distance (3.3\%) and integrated flux (13.5\%), this luminosity 
estimate, $L = 10^{4.41\pm0.06 }~L_{\odot}$, has a 15\% uncertainty, and it is also 15\% lower 
than our estimate $L = 10^{4.47\pm0.07}$  from photometry and bolometric correction.  Both 
luminosities agree within their propagated uncertainties using the relations 
$L = 4 \pi d^2 f = 4 \pi R^2 \sigma_{\rm SB} T_{\rm eff}^4$.   Our values, $T_{\rm eff} = 25,180$~K
and $L = 10^{4.41\pm0.06}$, are similar to the luminosity quoted in L.\ Fossati \etal\ (2015).  
{\bf Table 1} summarizes the stellar parameters from this work and previous papers.

\bigskip

\subsection{Stellar Atmospheres and EUV Fluxes}

\bigskip

For $\beta$~CMa, J.\ Cassinelli \etal\ (1996) discussed two sets of stellar atmosphere 
parameters ($T_{\rm eff}$ and $\log g$).  Following the method of  A.\ Code \etal\ (1996),
they used the observed total flux and angular diameter to find $T_{\rm eff} = 25,180$~K,
the same as ours (eq.\ [2]) using the same method.  However, their value of surface 
gravity ($\log g = 3.4$) is lower than our value of 3.7.   They also discussed possible lower 
temperatures, $T_{\rm eff} = 23,250$~K, with $\log g = 3.5$, that were in better agreement 
with their line-blanketed model atmospheres.  However, they also noted that the higher 
$T_{\rm eff}$ models gave better agreement with the 
observed EUV fluxes\footnote{Cassinelli \etal\ (1996) suggested that backwarming by the 
shocked stellar wind could boost the temperature in the upper atmosphere where the Lyman 
continuum is formed.  A similar non-LTE wind effect was proposed by F.\ Najarro \etal\ (1996), 
involving doppler shifts and velocity-induced changes in density that affect the escape of \HI\ 
and \HeI\ resonance lines and their ground-state populations.  Both mechanisms are sensitive 
to mass-loss rates in the range $\dot{M} \approx$ (1--10)$\times10^{-9}~M_{\odot}~{\rm yr}^{-1}$.}.
 At their larger adopted distance, $d = 206$~pc, the implied stellar parameters 
 ($R \approx 11.5~R_{\odot}$ and $M = gR^2/G \approx 24~M_{\odot}$) are discrepant with
 the derived position of $\beta$~CMa on the H-R diagram.   In a spectroscopic analysis of 
$\beta$~CMa, L.~Fosatti \etal\ (2015) found $T_{\rm eff} = 24,700\pm300$~K and 
 $\log g = 3.78\pm0.08$.  

\medskip

 For consistency  with the key observed parameters (distance, total flux, interferometric diameter), 
 we adopt $T_{\rm eff} = 25,180\pm1120$~K  and $\log g = 3.70\pm0.08$ 
 ($g = GM/R^2 = 5010\pm500~{\rm cm~s}^{-2}$).  With a measured rotational velocity  
 $V_{\rm rot} \sin i = 20.3\pm7.2$~\kms\ (L.\ Fosatti \etal\ 2015), we can neglect the centrifugal term, 
 $V_{\rm rot}^2/R \approx (7~{\rm cm~s}^{-2})(\sin i)^{-2}$ and infer a gravitational mass
 ($M = gR^2/G$) of  
\begin{equation}
   M = (13 \pm 3~M_{\odot}) \left( \frac {g}{5010~{\rm cm~s}^{-2}} \right) 
        \left( \frac {R} {8.44~R_{\odot}} \right)^2   \; .
\end{equation}
The quoted error on $M$ includes uncertainties in surface gravity and radius, added in quadrature 
with $(\sigma_M/M)^2 = (\sigma_g/g)^2 + (2 \sigma_R/R)^2$.   This gravitational mass is consistent 
with its evolutionary mass on the H-R diagram, $M_{\rm evol} = 13\pm1~M_{\odot}$. {\bf Figure 1} 
shows this location at $\log (L/L_{\odot}) = 4.41$ and $T_{\rm eff} = 25,180$~K, 
with evolutionary tracks from L.\ Brott \etal\ (2011).  This position is consistent with luminosity 
class~II/III (giant), and it places $\beta$~CMa within the $\beta$~Cephei instability strip, consistent 
with its observed pulsations. This conclusion is confirmed by comparison with the locus of radial and 
non-radial instability modes in Figs.\ 2 and 3  of L.\ Deng \& D.~R.\ Xiong (2001).

\bigskip

\section{\bf Interstellar Absorption and \HI\ Column Density } 

\bigskip

\subsection{Previous Column Density Estimates}

\bigskip

The EUVE observations of $\beta$~CMa (J.\ Cassinelli \etal\ 1996; Craig \etal\ 1997) showed
detectable flux between 510~\AA\ and 720~\AA, but no flux was detected below the \HeI\ edge 
($\lambda \leq 504.26$~\AA) or \HeII\ edge ($\lambda < 227.84$~\AA).  The lack of observed 
\HeI\ continuum flux is almost certainly the result of the large optical depth ($\tau \approx 3.35$
at 504~\AA)
arising from \HI\ and \HeI\ photoelectric absorption in the intervening ISM.  The two B-type stars, 
$\epsilon$~CMa  and $\beta$~CMa, are the strongest sources of EUV radiation as viewed 
from low-Earth orbit (J.\ Vallerga \& B.\ Welsh 1995).  This is primarily a result of their location
in a low-density cavity or  ``interstellar tunnel" (C.\ Gry \etal\ 1985; B.\ Welsh 1991; 
J.\ Vallerga \etal\ 1998).  In order to find the photoionizing flux incident on the exterior surface 
of the local clouds, we must correct for EUV flux attenuation by \HI\ and \HeI.  

\medskip

Previous estimates of the amount of interstellar hydrogen toward $\beta$~CMa were somewhat 
uncertain.  The {\it Copernicus} spectroscopic fits to the wings of the interstellar \Lya\ absorption 
line set upper limits of $N_{\rm HI} < 5\times10^{18}~{\rm cm}^{-2}$ (R.\ Bohlin 1975; 
R.\ Bohlin \etal\ 1978) and $N_{\rm HI} < 3\times10^{18}~{\rm cm}^{-2}$ 
(J. M.\ Shull \& M.\ Van Steenberg 1985).  The latter limit corresponds to an optical depth 
$\tau_{\rm LL} < 19$  at 912~\AA.  The actual ISM column density is less, since some EUV flux 
penetrates the local clouds.  The EUVE fluxes (510--720~\AA) detected by EUVE
(J.\ Cassinelli \etal\ 1996) were combined with stellar atmosphere models to estimate
an interstellar column density $N_{\rm HI} =$ (2.0--2.2)$\times10^{18}~{\rm cm}^{-2}$.  
In the next subsection, we make similar calculations and find a lower value,
$N_{\rm HI} = (1.9\pm0.1)\times10^{18}~{\rm cm}^{-2}$.  

\bigskip

\subsection{Photoelectric Absorption} 

\bigskip

In our previous study of $\epsilon$~CMa (J. M. Shull \etal\ 2025) we derived the intervening
\HI\ column density by comparing the observed EUV and FUV continuum fluxes from 
300--1150~\AA\ to non-LTE model atmospheres.  By exploring various attenuation models, 
we found an interstellar column density $N_{\rm HI} = (6\pm1)\times10^{17}$~cm$^{-2}$. 
For $\beta$~CMa, with less complete coverage of the EUV, we modified the method to
compare EUVE and model fluxes at four wavelengths (700, 650, 600, 570~\AA) and 
anchored to the observed fluxes at 1000--1200~\AA\ from Voyager and 
International Ultraviolet Explorer (IUE).  

\medskip

The expected FUV/EUV continua were based on model atmospheres produced with 
the code \texttt{WM-basic} developed by 
A.\ Pauldrach \etal\ (2001)\footnote{This code can be found at 
http://www.usm.uni-muenchen.de/people/adi/Programs/Programs.html}.
We chose this code because of its hydrodynamic solution of expanding atmospheres with 
line blanketing and non-LTE radiative transfer, including its treatment of the continuum and
wind-blanketing from  EUV lines.  Extensive discussion of hot-star atmosphere codes appears 
in papers by D.~J.\ Hillier \& D.\ Miller (1998), F.\ Martins \etal\ (2005), and C.\ Leitherer \etal\ (2014).  
We restore the observed continuum to its shape at the stellar surface by multiplying the observed 
flux by $\exp(\tau_{\lambda})$, using optical depths $\tau_{\lambda}$ of photoelectric absorption 
in the ionizing continua of \HI\  ($\lambda \leq 912$~\AA) and \HeI\ ($\lambda \leq 504$~\AA),
\begin{eqnarray}
 \tau_{\rm HI} (\lambda) &\approx& (6.304) \left( \frac {N_{\rm HI} } {10^{18}~{\rm cm}^{-2} } \right) 
           \left( \frac {\lambda} {912~{\rm \AA} } \right)^3   \; , \\
 \tau_{\rm HeI} (\lambda) &\approx& (0.737) \left( \frac {N_{\rm HeI} } {10^{17}~{\rm cm}^{-2} } \right) 
           \left(  \frac {\lambda} {504~{\rm \AA} } \right)^{1.63}   \; .   
\end{eqnarray}
These approximations are based on power-law fits to the photoionization cross sections at 
wavelengths below threshold,
$\sigma_{\rm HI}(\lambda) \approx (6.304\times10^{-18}~{\rm cm}^{2})(\lambda/912~{\rm \AA})^3$
(D.\ Osterbrock \& G.\ Ferland 2006) and 
$\sigma_{\rm HeI}(\lambda) \approx (7.37\times10^{-18}~{\rm cm}^{2})(\lambda/504~{\rm \AA})^{1.63}$
(D.\ Samson \etal\ 1994).  In our actual calculations, we used the exact (non-relativistic) \HI\ 
cross section (Bethe \& Salpeter 1957),
\begin{equation} 
   \sigma_{\nu} = \sigma_0 \left( \frac {\nu}{\nu_0} \right)^{-4} 
             \frac {\exp[ 4 - (4 \arctan \epsilon ) / \epsilon ]}  { [1 - \exp(-2 \pi / \epsilon) ] }  \; .
\end{equation}
Here, the dimensionless parameter $\epsilon \equiv [(\nu/\nu_0) -1]^{1/2}$ with frequency $\nu_0$ 
defined at the ionization energy $h \nu_0 = 13.598$~eV and 
$\sigma_0 =  6.304\times10^{-18}~{\rm cm}^{2}$.  The two formulae agree at threshold $\nu = \nu_0$, 
but the approximate formula deviates increasingly at shorter wavelengths.  The exact cross section is 
higher by 8.2\% (700~\AA), 12.3\% (600~\AA), and 16.4\% (500~\AA).  
 
 \medskip
   
In the far-UV, the stellar continuum in the non-LTE model atmosphere rises slowly from 1150~\AA\ 
down to 1000~\AA, and then declines owing to absorption in higher Lyman-series lines converging 
on the LL at 911.75~\AA.  For reference, a pure blackbody at $T = 25,180$~K peaks at 
$\lambda_{\rm max} \approx 1160$~\AA.  However, there is still some uncertainty in the spectral 
shape from 1150~\AA\ down to 912~\AA, a region unobserved by IUE.  The Voyager observations, 
plotted in Figure 2a of J.\ Cassinelli \etal\ (1996), show a peak in 
$F_{\lambda} \approx 4\times10^{-8}~{\rm erg~cm}^{-2}~{\rm s}^{-1}~{\rm \AA}^{-1}$ around 
1000~\AA, similar to the mean flux seen in IUE (SWP Large Aperture) archival spectra at
1150~\AA.   

\medskip

 For our flux attenuation calculations, we adopt a continuum level at $912^+$~\AA, just longward 
of the Lyman edge, of $F_{\lambda} = 4\times10^{-8}~{\rm erg~cm}^{-2}~{\rm s}^{-1}~{\rm \AA}^{-1}$.
This corresponds to a photon flux of
 $\Phi(912^+) = 1836$ photons~cm$^{-2}~{\rm s}^{-1}~{\rm \AA}^{-1}$.  In the  \texttt{WM-basic} 
 model atmosphere {\bf (Figure 2)}, the stellar flux drops by a factor $F(912^+)/F(912^-) = 67\pm7$.  
We then compare the model fluxes to the photon fluxes $\Phi(\lambda) = \lambda F_{\lambda}/hc$ 
at four EUVE-observed wavelengths, with photon fluxes (photons cm$^{-2}$~s$^{-1}$~\AA$^{-1}$) 
of 0.02 (700~\AA), 0.05 (650~\AA), 0.085 (600~\AA), and 0.11 (570~\AA).  After converting 
$F_{\lambda}$ to $\Phi_{\lambda}$, we predict model photon flux ratios of
 \begin{eqnarray}
    \Delta(700) &=& \Phi(912^+)/\Phi(700) = 209  \\
    \Delta(650) &=& \Phi(912^+)/\Phi(650) = 468  \\
    \Delta(600) &=& \Phi(912^+)/\Phi(600) = 788  \\
    \Delta(570) &=& \Phi(912^+)/\Phi(570) = 1179   \;  .
 \end{eqnarray} 
 Combining the observed photon fluxes, model flux ratios, and anchor flux $\Phi(912^+)$, we 
 arrive at estimates of the ISM optical depth,
 \begin{equation}
    \tau_{\rm ISM} = \ln \left[ \frac {\Phi(912^+)/\Phi(\lambda)} {\Delta(\lambda)} \right]   \; ,
 \end{equation} 
 with a corresponding \HI\ column density $N_{\rm HI} = \tau_{\rm ISM} / \sigma_{\rm HI} (\lambda)$.  
 The most reliable estimates come from data and models at 
 700~\AA\ ($\sigma_{\rm HI} = 3.086\times10^{-18}~{\rm cm}^{2}$) and
 650~\AA\ ($\sigma_{\rm HI} = 2.516\times10^{-18}~{\rm cm}^{2}$), for which we obtain:
 \begin{eqnarray}
     \tau_{\rm ISM}(700~{\rm \AA}) &=&  5.964 \; \; {\rm and} \; \; 
            N_{\rm HI} = 1.93\times10^{18}~{\rm cm}^{-2}   \\
     \tau_{\rm ISM}(650~{\rm \AA}) &=&  4.608 \; \; {\rm and} \; \; 
            N_{\rm HI} = 1.83\times10^{18}~{\rm cm}^{-2}  \;  .
 \end{eqnarray}  
 We adopt a mean value $N_{\rm HI} = (1.9\pm0.1)\times10^{18}~{\rm cm}^{-2}$, with 
 optical depth $\tau_{\rm ISM} = 12.0 \pm 0.6$ at the Lyman limit.  The stellar atmosphere
 produces a LL flux decrement factor $\Delta_{\rm star} \approx 67\pm7$, and the ISM produces 
 $\Delta_{\rm ISM} \approx e^{12} \approx 1.6\times 10^5$. With this much attenuation, there
 would be little detectable flux from 750~\AA\ to 912~\AA.  The EUVE spectrometers did not observe 
 at wavelengths longward of 730~\AA, and the Colorado Dual-Channel Extreme Ultraviolet Continuum 
 Experiment (DEUCE) spectrograph saw no flux below 912~\AA\ in its (2 November 2020) rocket flight 
 (Emily Witt, private communication).  
 
 \bigskip

\subsection{Photoionization Rates Outside the Local Clouds}

\bigskip

With our derived interstellar column density toward $\beta$~CMa of
$N_{\rm HI} = (1.9\pm0.1)\times10^{18}~{\rm cm}^{-2}$, we adopt a \HeI\ column density 
$N_{\rm HeI} = (1.3\pm0.2)\times10^{17}~{\rm cm}^{-2}$, based on observations of interstellar
ratios of \HeI/\HI\ toward nearby white dwarfs\footnote{J.\ Dupuis \etal\ (1995) reported a mean 
ratio $\beta = N_{\rm HI} / N_{\rm HeI}  \approx 14$, after correcting the observed \HI\ and \HeI\ 
column densities for stellar contributions using  LTE, hydrostatic, plane-parallel model atmospheres.  
The EUVE  column densities indicated $\beta$ ranging from $12.1\pm3.6$ to $18.4\pm1.3$.  
Removing the two extreme values, we obtain a mean and rms dispersion of $\beta = 14.5\pm1.1$.
These values are consistent with earlier sounding-rocket results (J.\ Green \etal\ 1990) with
$11.5 < \beta < 24$ (90\% confidence level) toward the white dwarf G191-B2B.}.  
The photoionization rates of \HI\ and \HeI\ are found by integrating the ionizing photon flux
$\Phi_{\lambda} =  (\lambda \, F_{\lambda} / hc)$ multiplied by the photoionization cross section
$\sigma_{\lambda}$ and the attenuation factor $\exp (\tau_{\lambda})$.  
For \HI\ ionization, we use the detected EUVE fluxes (504--730~\AA) and an extrapolation of
the product $\Phi_{\lambda} \sigma_{\lambda} \exp (\tau_{\lambda})$ from 730~\AA\ to 912~\AA.
After correcting for ISM attenuation, we integrate over the ionizing spectrum to find a total photon 
flux of $\Phi_{\rm H} = 3700\pm1000~{\rm cm}^{-2}~{\rm s}^{-1}$ and an \HI\ photoionization rate 
$\Gamma_{\rm HI} = 1.5\times10^{-14}~{\rm s}^{-1}$ at the external surface of the local cloud.  
These values are comparable to those of $\epsilon$~CMa (Shull \etal\ 2025), with 
$\Phi_{\rm H} = 3100\pm1000~{\rm cm}^{-2}~{\rm s}^{-1}$ and 
$\Gamma_{\rm HI} = 1.3\times10^{-14}~{\rm s}^{-1}$.

\medskip

For the \HeI\ continuum, we again use the non-LTE model atmosphere  {\bf (Figure 2)} to establish 
the flux ratio between the two ionization edges, $F(912^+)/F(504^+) \approx 7000$.  The flux falls off 
rapidly below the \HeI\  threshold and is well fitted (between $400~{\rm \AA} \leq \lambda \leq 504$~\AA) 
by the relation $F_{\lambda} = F_0 (\lambda/\lambda_0)^{10}$, with 
$F_0 = 5.7\times10^{-12}~{\rm erg~cm}^{-2}~{\rm s}^{-1}~{\rm \AA}^{-1}$ at 
$\lambda_0 = 504.26$~\AA. The \HeI\ cross section is
$\sigma_{\rm HeI} \approx \sigma_0 (\lambda/\lambda_0)^{1.63}$, where 
$\sigma_0 = 7.37\times10^{-18}~{\rm cm}^2$ at 504.26~\AA. We integrate over wavelengths 
$\lambda \leq \lambda_0$, defining a dimensionless variable $u = (\lambda / \lambda_0)$ and 
expressing the surface photoionization integral as
\begin{equation}
   \Gamma_{\rm HeI} = \int_{0}^{\lambda_0}  \left[ \frac {\lambda \, F_{\lambda}}
         {hc} \right] \, \sigma_{\rm HeI} (\lambda) \,  e^{\tau(\lambda)} \, d \lambda  
   = \left[ \frac {F_0 \, \sigma_0 \, \lambda_0^2} {hc} \right] \int_{0}^{1} u^{12.63} \, e^{\tau(u)}  du \;  .
\end{equation} 
Similarly, the total flux in the \HeI\ ionizing continuum can be expressed as,
\begin{equation}
   \Phi_{\rm HeI} = \int_{0}^{\lambda_0}  \left[ \frac {\lambda \, F_{\lambda} }
         {hc} \right] \;  e^{\tau(\lambda)} \,  d \lambda  
   = \left[ \frac {F_0 \, \lambda_0^2} {hc} \right] \int_{0}^{1} u^{11} \, e^{\tau(u)}   du \;  .
\end{equation} 
The two dimensionless integrals are 1.35 (eq.\ [15]) and 1.46 (eq.\ [16]), showing increases by
factors of 18 (mean $\tau = 2.86$) over the unattenuated integrals (setting $e^{\tau} = 1$).  We 
then obtain $\Gamma_{\rm HeI} = 7.3\times10^{-16}~{\rm s}^{-1}$ and 
$\Phi_{\rm HeI} = 107$ photons cm$^{-2}$~s$^{-1}$.  Because of its hotter effective temperature, 
$\beta$~CMa is an important source of He-ionizing photons in the local cloud, much larger than 
$\epsilon$~CMa for which we estimate $\Gamma_{\rm HeI} = 4.4\times10^{-17}$~s$^{-1}$.   
As noted by  J.\ Dupuis \etal\ (1995) and J.\ Vallerga (1998), other sources of \HeI\ ionization include
 three nearby white dwarfs (G191-B2B, Feige~24, HZ~43A).

\medskip

The hydrogen and helium ionization fractions at the outer surface of the local cloud depend on 
the ionizing fluxes from two B-type stars ($\epsilon$~CMa and $\beta$~CMa), three white dwarfs,
and EUV emission lines produced by hot plasma in the local hot bubble (LHB).  The 
two B-type stars produce a combined LyC photon flux (at the external cloud surface) of 
$\Phi_{\rm HI} \approx 6800\pm1400~{\rm cm}^{-2}~{\rm s}^{-1}$ and hydrogen ionization
rate $\Gamma_{\rm HI}  \approx 3\times10^{-14}~{\rm s}^{-1}$.  In photoionization equilibrium, 
the hydrogen ionization fraction, $x = n_{\rm HII}/n_{\rm H}$, is $x = (1/2)[ -a + (a^2 + 4a)^{1/2}]$,  
the solution of $x^2/(1-x) = a$, where the constant 
$a = [\Gamma_{\rm HI} / 1.1 n_{\rm H} \alpha_{\rm H}]$.  We adopt a hydrogen case-B 
radiative recombination coefficient (Draine 2011) 
$\alpha_{\rm H} = 3.39\times10^{-13}~{\rm cm}^3~{\rm s}^{-1}$ at $T = 7000$~K. 
The cloud is assumed to have constant hydrogen density 
$n_{\rm H} \equiv n_{\rm HI} + n_{\rm HII} \approx 0.2~{\rm cm}^{-3}$, and the factor 1.1 
accounts for electrons potentially contributed by He$^+$.  We find $a = 0.402$ and 
$x = 0.464$ at the surface of the local clouds.  If we include the He-ionizing continuum, with
$\Gamma_{\rm HeI} = 7.3\times10^{-16}$~s$^{-1}$ and solve coupled equations for the 
H and He ionization equilibrium, we find $x \approx 0.479$ (H$^+$) and $y \approx 0.022$ (He$^+$). 
These ionization fractions decrease with depth into the cloud, as the LyC flux is attenuated. 

\bigskip

\subsection {Stellar motion and Non-equilibrium Ionization}  

\bigskip

For $\beta$~CMa, Hipparcos measurements (F.\ van Leeuwen 2007) find a parallax distance 
$d = 151\pm5$~pc,  radial velocity $V_r = 33.7\pm0.5$~\kms, and transverse velocity 
$V_{\perp} = 2.38\pm0.21$~\kms.  The transverse velocity is based on its total proper motion 
$\mu_{\perp} = -3.32\pm0.28~{\rm mas~yr}^{-1}$ with components   
$\mu_{\alpha} \,\cos \delta  = -3.23\pm0.19~{\rm mas~yr}^{-1}$ and 
$\mu_{\delta} = -0.78\pm0.20~{\rm mas~yr}^{-1}$ (in RA and Decl).  
Thus, $\beta$~CMa passed close to the Sun at a time 
$t_{\rm pass} = d/V_r = 4.38\pm0.16$~Myr ago, at an offset distance of 
$d_{\perp} = \mu_{\perp} d^2/V_r = 10.6\pm1.1$~pc.  These values are similar to those
for $\epsilon$~CMa, which passed by the Sun $4.44\pm0.10$~Myr at offset distance 
$9.3\pm0.5$~pc (Shull \etal\ 2025).   The ionizing radiation field from these stars would have
been 200 times higher than the current value.  
With the Sun's trajectory moving out of the local cloud, the location of $\epsilon$~CMa and
$\beta$~CMa relative to the local hydrogen gas remains uncertain.  The local clouds are not
self-gravitating and probably gravitationally unbound from the Sun.  They may not have been
exposed to the same enhanced ionizing radiation field.  

\bigskip

Nevertheless,  $\epsilon$~CMa and $\beta$~CMa would have produced strong ionizing effects 
in the past on any nearby interstellar gas. The stellar radial velocities correspond to distance changes 
of 27.3~pc/Myr ($\epsilon$~CMa) and 33.7~pc/Mpc ($\beta$~CMa) away from the Sun.  Tracking the 
enhancements in $\Gamma_{\rm H}(t)$ along the past trajectories of these stars, we estimate that the 
hydrogen ionization fraction peaked at $x \approx 0.99$ at closest passage (4.4 Myr ago) and the LIC
would have been fully ionized.  At 1.6 Myr in the past, when the stars were at distances of 79~pc 
($\epsilon$~CMa) and 96~pc ($\beta$~CMa), the equilibrium ionization fraction decreased to 
$x \approx 0.62$. More recently, the fraction declined to values $x \approx 0.5$.  Over the last
1.6~Myr, non-equilibrium ionization effects set in, as the recombination time becomes longer than 
the photoionization time.  

\medskip

To illustrate this effect, we estimate the time scales for hydrogen photoionization and recombination,
and for radiative cooling of gas near the local cloud surface:
\begin{eqnarray}
    t_{\rm ph}    &=& \Gamma_{\rm H}^{-1} \approx (1~{\rm Myr}) 
                              \left( \frac {\Gamma_{\rm H}}{3\times10^{-14}~{\rm s}^{-1}}  \right)^{-1} \\
    t_{\rm rec}  &=& (n_e \alpha_{\rm H})^{-1}  \approx (0.93~{\rm Myr}) 
                             \left( \frac {n_e}{0.1~{\rm cm}^{-3}} \right)^{-1} T_{7000}^{0.809}  \\
    t_{\rm cool} &=&  \frac {3n_{\rm tot} kT/2} { n_{\rm H}^2 \, \Lambda(T)} 
                          \approx (11~{\rm Myr}) \left( \frac {n_{\rm H}}{0.2~{\rm cm}^{-3}} \right)^{-1}  
                           T_{7000}    \; . 
\end{eqnarray}
Here, we scaled $\Gamma_{\rm H} $ to the combined rates for $\beta$ and $\epsilon$~CMa
at the LIC and used a case-B radiative recombination rate coefficient 
$\alpha_{\rm H} = (3.39\times10^{-13}~{\rm cm}^3~{\rm s}^{-1})T_{7000}^{-0.809}$ (B.\ Draine 2011),
appropriate for $T = (7000~{\rm K})T_{7000}$.  We adopted total hydrogen density 
$n_{\rm H} \approx 0.2$~cm$^{-3}$, electron density $n_e \approx 0.1~{\rm cm}^{-3}$, and a 
radiative cooling rate $n_{\rm H}^2 \Lambda(T)$ with coefficient 
$\Lambda(T) \approx 3\times10^{-26}~{\rm erg~cm}^3~{\rm s}^{-1}$. 

\medskip

The current timescales for photoionization and recombination are comparable ($\sim 1$~Myr), but 
the radiative cooling time and stellar passing time are considerably longer. Any gas clouds in the 
Sun's vicinity 4 Myr ago would have been highly ionized.   These two CMa stars left a wake of 
ionized and heated gas, consistent with the tunnel of low N$_{\rm HI}$ observed in their direction.  
A high level of ionization was established, peaking at $x \approx 0.99$ during closest stellar passage.
The ionization levels of H and He decreased on Myr timescales, tracking the motions of the stars, with 
non-equilibrium effects setting in during the last 1.6 Myr.  In the future, the Sun will exit the local cloud and 
once again be exposed to a much higher ionizing radiation field unshielded by the local cloud.

\bigskip

\section{Summary of Results and Future Studies}  

\bigskip

We derived a consistent set of stellar parameters for $\beta$~CMa (mass, radius, effective 
temperature, luminosity) consistent with its shorter parallax distance (151~pc vs.\ 203~pc), 
interferometric angular diameter ($\theta_d = 0.52\pm0.03$~mas), and integrated bolometric 
flux, $f = (36.2 \pm 4.9) \times10^{-6}~{\rm erg~cm}^{-2}~{\rm s}^{-1}$.  From these, we derived
$T_{\rm eff} = 25,180\pm1120$~K and $L \approx 10^{4.41\pm0.06}~L_{\odot}$ and updated values of
absolute magnitude ($M_V = -3.93\pm0.04$ and $M_{\rm bol} = -5.97$).  On the H-R diagram, 
the new parameters place $\beta$~CMa within the $\beta$-Cephei pulsational instability strip, 
consistent with its observed pulsations and the boundaries of theoretical instability on evolutionary 
tracks (L.\ Deng \& D.~R.\ Xiong 2001).  The following points summarize our primary results:
\begin{enumerate}

\item The stellar parallax distance and angular diameter of $\beta$~CMa yield a radius 
$R = 8.44\pm0.56~R_{\odot}$.  Bolometric relations between the integrated stellar flux and 
radius yield an effective temperature $T_{\rm eff} \approx 25,180\pm1120$~K and luminosity 
$L \approx 10^{4.41\pm0.06}~L_{\odot}$.  Both gravitational and evolutionary masses are 
consistent at $M \approx 13\pm1~M_{\odot}$.  

\item From models of the stellar and interstellar attenuation of the ionizing flux in the 
Lyman continuum, we determine a column density 
$N_{\rm HI} = (1.9\pm0.1) \times 10^{18}~{\rm cm}^{-2}$ in the local cloud,
corresponding to optical depth $\tau_{\rm LL} = 12.0\pm0.6$ at the Lyman limit.   

\item Using non-LTE model atmospheres and observed EUV spectra, we estimate a stellar
flux decrement $\Delta_{\rm star} = 67\pm7$ at the Lyman limit.  At the external surface of 
the local cloud, $\beta$~CMa produces ionizing photon fluxes of 
$\Phi_{\rm HI} \approx 3700\pm1000~{\rm cm}^{-2}~{\rm s}^{-1}$ and
$\Phi_{\rm HeI} \approx 110 \pm 30~{\rm cm}^{-2}~{\rm s}^{-1}$.  At a distance 
$d = 151\pm5$~pc, its LyC  photon production rate is 
$Q_{\rm H} \approx 10^{46.0\pm0.1}~{\rm photons~s}^{-1}$. 
 
\item  Because of attenuation, the EUV flux from $\beta$~CMa incident on the local clouds 
is $\sim20$ times higher than viewed from Earth.  The photoionization rates at the cloud
surface are $\Gamma_{\rm HI} \approx 1.5 \times 10^{-14}~{\rm s}^{-1}$ and 
$\Gamma_{\rm HeI} \approx 7.3\times10^{-16}$~s$^{-1}$.  For local gas with 
$n_{\rm H} = 0.2~{\rm cm}^{-3}$ and $T \approx 7000$~K, photoionized only by
$\epsilon$ and $\beta$~CMa, the current ionization fractions at the cloud surface would
be  $x \approx 0.48$ (H$^+$) and $y \approx 0.02$ (He$^+$).  They would have been much 
higher in the past (4.4 Myr ago) when both $\epsilon$~CMa and $\beta$~CMa passed 
within $10\pm1$~pc of the Sun 
.  

\end{enumerate} 

\medskip 

Several ionization issues remain, particularly the potentially enhanced level of He$^+$ ionization.
Additional sources of He-ionizing photons could arise from three nearby white dwarfs or from the 
local hot bubble, a cavity of hot plasma ($T \approx 10^6$~K) believed responsible for the soft X-ray 
background (Snowden \etal\ 1990; Galeazzi \etal\ 2014; Yueng \etal\ 2024).  We are exploring such 
models in a subsequent paper, where we find that the hot bubble could produce ionizing photon fluxes 
$\Phi_{\rm H} =$7000--9000~cm$^{-2}$~s$^{-1}$, comparable to or greater than those of the B stars.
Importantly, this emission includes flux in the \HeI\ continuum (100--500~\AA) produced by a range 
of ionization states of iron (Fe$^{+8}$ to Fe$^{+12}$.  Recent reviews of the local ISM 
(J.\ Slavin \& P.\ Frisch 2008;  P. Frisch \etal\ 2011) suggest mean ionization fractions $x =$ 0.2--0.3 
and $y =$ 0.4--0.5, and electron fractions $f_e =$ 0.24--0.35.  The elevated He$^+$ fractions would 
require $\Gamma_{\rm HeI} \approx$ (2--3)$\Gamma_{\rm HI}$ in photoionization equilibrium conditions.  
As suggested by Slavin \& Frisch (2002), the EUV emission from various Fe ions in the hot plasma 
may help to explain the elevated He$^+$/H$^+$ ionization ratios.   Alternatively, the He$^+$ and 
H$^+$ could be out of photoionization equilibrium, as a result of the trajectories of $\epsilon$~CMa 
and $\beta$~CMa relative to the Sun over the past 4.4~Myr.

\begin{acknowledgements}

 We thank Emily Witt, James Green, and Kevin France for discussions of the far-UV 
 spectra of $\beta$~CMa with the Colorado DEUCE rocket.   
 We also thank Jeffrey Linsky, Seth Redfield, and Jon Slavin for scientific discussions 
 about the local ISM.  A portion of this study was supported by the New Horizons Mission 
 program for astrophysical studies of cosmic optical, ultraviolet, and \Lya\ backgrounds.

\end{acknowledgements} 

%   \newpage

%%%%%%%%%%%%%%%   REFERENCES   %%%%%%%%%%%%%%%%%%

%%%%%%%%%%%%%%%%%%%%%%%%%%%%%%%%%%%%%%%%%%

%    \clearpage

%%%%%%%%%%%%%%%%%%%%%%%%%%%%%%%%%%%%%%%%%%%%%
%%
%%                      Table 1 (Stellar Parameters)                                     %%%%%%
%%
%%%%%%%%%%%%%%%%%%%%%%%%%%%%%%%%%%%%%%%%%%%%%

\begin{deluxetable} {lccc cccc}
\tablecolumns{8}
 \tabletypesize{\scriptsize}

\tablenum{1}
\tablewidth{0pt}
\tablecaption{Various Stellar ($\beta$~CMa) Parameters\tablenotemark{a} }  

\tablehead{
   \colhead{Reference Paper}
 & \colhead{$d$}
 & \colhead{$T_{\rm eff}$} 
 & \colhead{$\log g$} 
 & \colhead{$R/R_{\odot}$}
 & \colhead{$M/M_{\odot}$} 
 & \colhead{$L/L_{\odot}$} 
 & \colhead{$M_{\rm bol}$}
 \\
 \colhead{}
 & \colhead{(pc)}
 & \colhead{(K)} 
 & \colhead{(cgs)} 
 & \colhead{}
 & \colhead{} 
 & \colhead{} 
 & \colhead{(mag)} 
 }

\startdata
Cassinelli \etal\ (1996)  & 206 & $20,990\pm760$ & $3.4\pm0.15$ & $16.2^{+1.2}_{-1.2}$  
                        & $15.2^{+6.4}_{-4.4}$ & $45,900\pm9500$ & $-6.91$  \\
Fossati \etal\ (2015)\tablenotemark{b} & $151\pm5$ & $24,700\pm300$ & $3.78\pm0.08$ & $7.4^{+0.8}_{-0.9}$ 
                         & $12.0^{+0.3}_{-0.7}$ & $25,700^{+3800}_{-3800}$ & $-6.29$ \\ 
Fossati \etal\ (2015)\tablenotemark{c}  & $151\pm5$ & $24,700\pm300$ & $3.78\pm0.08$ & $8.2^{+0.6}_{-0.5}$ 
                         & $12.6^{+0.4}_{-0.5}$ & $25,700^{+3800}_{-3300}$ & $-6.29$  \\ 
Current Study (2025) & $151\pm5$  & $25,180\pm1120$ & $3.70\pm0.08$ & $8.44\pm0.57$ & $13\pm1$ 
                         & $25,800\pm3900$  & $-5.97$    \\
\enddata 

\tablenotetext{a} {Values of effective temperature, surface gravity, radius, mass, luminosity,
and bolometric absolute magnitude given in past papers.  L.\ Fossati \etal\ (2015) derived
$R$ and $M$ from two sets of evolutionary tracks (footnotes b and c). }

\tablenotetext{b} {Stellar mass and radius inferred from evolutionary tracks of C.\ Georgy \etal\ (2013). }
\tablenotetext{c} {Stellar mass and radius inferred from evolutionary tracks of L.\ Brott \etal\ (2011). }

\end{deluxetable}

%%%%%%%%%%%%%%%%%%%%%%%%%%%%%%%%%%%%%%%%%%%%%%%%%

\newpage

%%%%%%%%%%%%%%%%%%%%%%%%%%%%%%%%%%%%%%%%%%
%%
%%               Figure 1 (HR Diagram:  Beta CMa and Eps CMa)
%%    
%%%%%%%%%%%%%%%%%%%%%%%%%%%%%%%%%%%%%%%%%%%

\begin{figure}[ht]
\centering
\includegraphics[angle=0,scale=0.8] {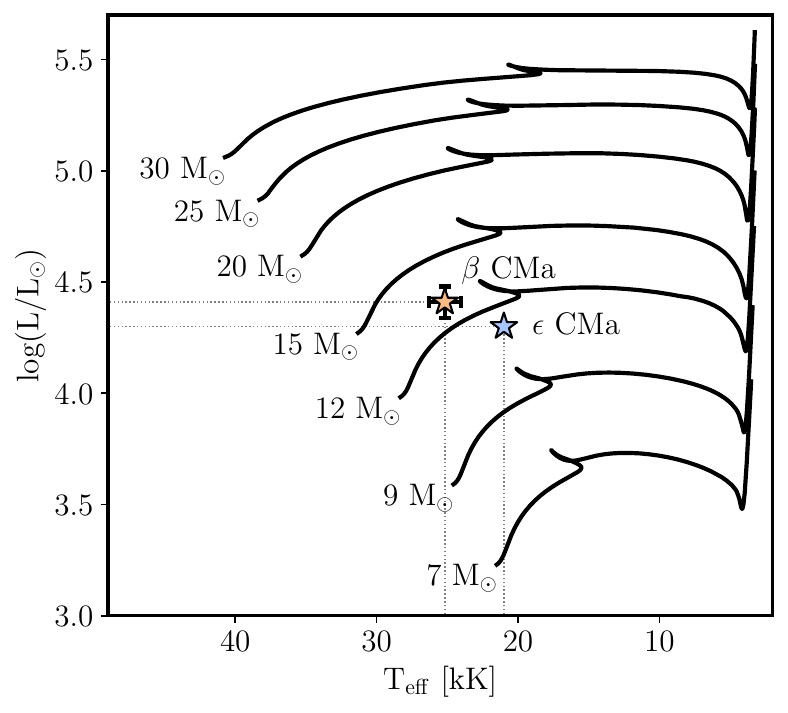}
\caption{The location of $\beta$~CMa on the Hertzsprung-Russell diagram is shown for 
our derived parameters, $\log (L/L_{\odot}) = 4.41\pm0.06$ and 
$T_{\rm eff} = 25,180\pm1120$~K, based on new radius $R = 8.44\pm0.56~R_{\odot}$ 
and parallax distance $d = 151\pm5$~pc.  The evolutionary tracks are from L.\ Brott \etal\ (2011) 
with Milky Way metallicities and initial masses labeled from 7--30 $M_{\odot}$.  
The location of $\epsilon$~CMa (J. M.\ Shull \etal\ 2025) is shown for comparison.
 }
\end{figure}
%%%%%%%%%%%%%%%%%%%%%%%%%%%%%%%%%%%%%%%%%%%

%%%%%%%%%%%%%%%%%%%%%%%%%%%%%%%%%%%%%%%%%%
%%
%%               Figure 2 (Model Atmospheres for Beta~CMa)
%%    
%%%%%%%%%%%%%%%%%%%%%%%%%%%%%%%%%%%%%%%%%%%

\begin{figure}[ht]
\includegraphics[angle=0,scale=0.59] {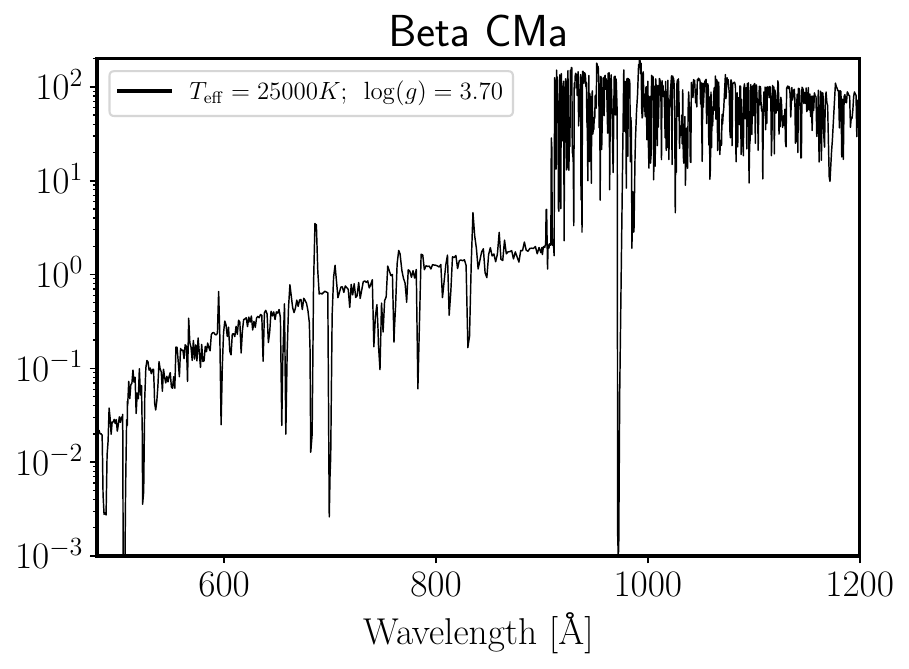}
\includegraphics[angle=0,scale=0.55] {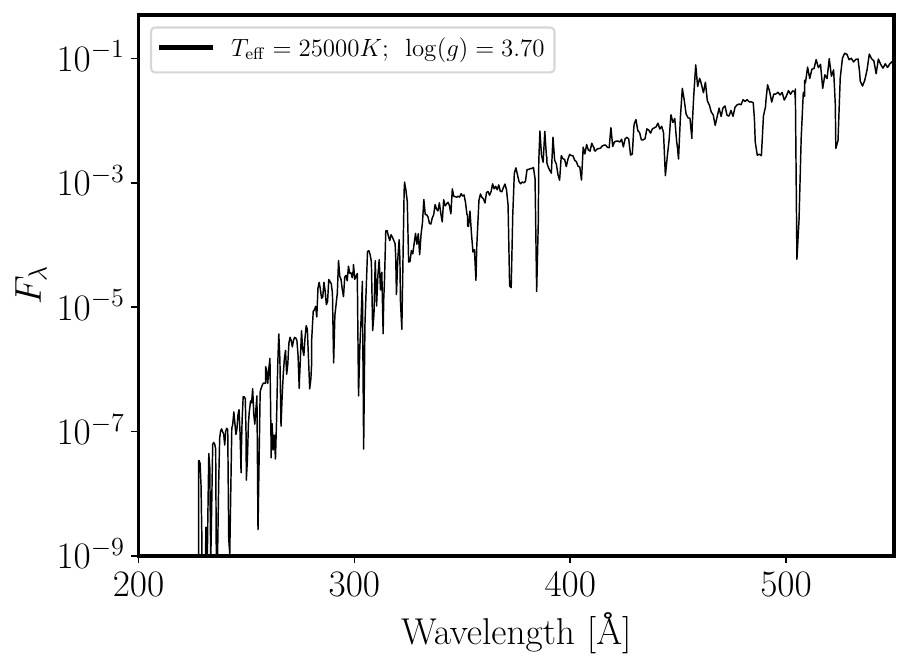}
\caption{Far-UV and EUV spectra for $\beta$~CMa from a model atmosphere
computed with the non-LTE line-blanketed code \texttt{WM-basic} for effective 
temperature $T_{\rm eff} = 25,000$~K  and surface gravity $\log g = 3.70$. 
(Left)   Flux distribution $\log F_{\lambda}$ from 850--1200~\AA, showing the 
Lyman limit decrement at 912~\AA.  
(Right)  Flux distribution from 228~\AA\ to 550~\AA.  The absence of an edge
at the \HeI\ ionization limit (504~\AA) is a result of non-LTE effects from 
backwarming of the upper atmosphere from a solar wind in early B-type stars.  
}

\end{figure}

%%%%%%%%%%%%%%%%%%%%%%%%%%%%%%%%%%%%%%%%%%%

\end{document}